\RequirePackage{fix-cm}

\documentclass{svjour3}                     
\smartqed 
\usepackage{subfigure}
\usepackage{bm}
\usepackage{dcolumn}
\usepackage{latexsym}
\usepackage{amsmath}
\usepackage{amsfonts}
\usepackage{amssymb}
\usepackage{theorem}
\usepackage{makeidx}
\usepackage{hyperref}
\usepackage{graphicx}
\usepackage{latexsym}
\usepackage{cite}
\usepackage{enumerate}
\usepackage[mathscr]{eucal}
\usepackage[usenames,dvipsnames]{color}
\usepackage{epstopdf}

\DeclareMathOperator{\id}{id}

\DeclareMathOperator{\kummerM}{M}

\newcommand{\D}{\frac{\mathrm{d}}{\mathrm{d}\tau}}

\newcommand{\Prob}{\mathbf{P}}

\newcommand{\dd}{\mathrm{d}}
\newcommand{\ee}{\mathrm{e}}
\newcommand{\transp}{\mathrm{T}}

\newcommand{\ignore}[1]{}

\journalname{Bulletin of Mathematical Biology}

\begin{document}

\DeclareGraphicsExtensions{.eps}

\title{Protein synthesis driven by dynamical stochastic transcription}

\author{\mbox{Guilherme C.P.\ Innocentini} \and
        \mbox{Michael Forger} \and
        \mbox{Ovidiu Radulescu} \and
        \mbox{Fernando Antoneli}}
        
\institute{Departamento de Matem\'atica Aplicada \\
           Instituto de Matem\'atica e Estat\'{\i}stica -
           Universidade de São Paulo \at
           Rua do Matão, 1010 - Cidade Universitária -
           São Paulo - SP - Brazil - CEP: 05508-090 \\
           \email{ginnocentini@gmail.com}\\
           \email{forger@ime.usp.br}
           \and
           DIMNP - UMR 5235 - Université de Montpellier 2  \at
           Pl.\ E.\ Bataillon - Bat. 24 - 34095 Montpellier Cedex 5 - France \\
	   \email{ginnocentini@gmail.com}\\ 
           \email{ovidiu.radulescu@univ-montp2.fr}
	   \and
           Laborat\'orio de Gen\^omica Evolutiva e Biocomplexidade \& DIS \\
           Escola Paulista de Medicina --
           Universidade Federal de S\~ao Paulo \at 
           Rua Pedro de Toledo, 669, 4th floor, 
           S\~ao Paulo - SP - Brazil - CEP: 04039-032 \\
           \email{fernando.antoneli@unifesp.br}}       

\date{\today}

\maketitle

\begin{abstract}
 In this manuscript we propose a mathematical framework to couple transcription
 and translation in which mRNA production is described by a set of master
 equations while the dynamics of protein density is governed by a random
 differential equation.
 The coupling between the two processes is given by a stochastic
 perturbation whose statistics satisfies the master equations.
 In this approach, from the knowledge of the analytical time
 dependent distribution of mRNA number, we are able to calculate
 the dynamics of the probability density of the protein population.

\keywords{Gene expression \and Stochasticity \and Exact solutions \and Dynamics}
\PACS{87.16.Ac, 87.10.+e, 87.16.Yc}
\subclass{92B05} 
\end{abstract}

\section{Introduction}
\indent

Stochasticity in biological processes, in particular of gene expression,
has been studied, both experimentally and theoretically, at least since the
pioneering work of Delbr\"uck \cite{delbruck}.
Recent advances in experimental methods have enabled direct observation
of stochastic features of gene expression, such as temporal fluctuations
in individual cells or steady-state variations across a cell population~%
\cite{Elowitz2002,OTKGO2002,BKCC2003,ROS2004,GPZC2005,CFX2006,YXRLX2006},
and data acquisition has experienced a huge improvement in the last decade.
However, theoretical models have not yet been developed to the point of
providing a comprehensive quantitative description for the dynamics of
gene expression. The stationary regime has been exhaustively discussed
in the literature, but studies on time dependent probability distributions
are still scarce~\cite{Jayaprakash2009,Swain2008,Ramos2011}.

In this paper, our main goal is to present and discuss a stochastic
description for mRNA-protein dynamics. More precisely, we propose and 
solve a hybrid model for stochastic gene expression, consisting of a master 
equation (ME) coupled to a random differential equation (RDE). The ME 
describes the production of messenger RNA (mRNA) molecules triggered by 
a gene with various levels of promoter activity. The RDE governs the 
dynamics of protein synthesis: it is a linear ordinary differential equation
randomly perturbed by the Markov jump process underlying the ME.
The master equation part of the model is a particular case of a Markov process
in a ``random environment''~\cite{CT1981}, composed by a birth-and-death 
process and a two-state markovian switching process, in 
continuous time; see~\cite{PY1995} for the interpretation in the context of 
gene expression.
Several variations of this type of model have been employed for the
study of gene expression and have been extensively discussed in the
literature~\cite{KE2001,PE2004,Hornos2005,Paulsson2005}.
The particular form of the master equation part used in this paper is the
one analyzed in~\cite{Innocentini2007,Innocentini2013}.

The motivations for such an approach can be justified
on mathematical as well as biological grounds.
From a mathematical point of view, the RDE employed here resembles a
Langevin equation, with one crucial difference: the driving stochastic
process is not a singular \emph{delta}-like noise, but rather a non-%
singular, well behaved stationary stochastic process.
Non-white noise driven Langevin-like equations have been widely
discussed in the literature under different names, such as \emph%
{colored noise}~\cite{vKampen2007} or \emph{real noise}~\cite{Arnold1998}. 
And the mathematical advantage in dealing with RDEs is that one does
not need a sophisticated theory of integration in order to solve them. 
As a matter of fact, RDEs are solved by Riemann integration of ordinary 
differential equations, sample path by sample path~-- hence the term
``random differential equation'' instead of the more familiar term
``stochastic differential equation'', which is reserved for
differential equations associated to a stochastic integration
theory~\cite{Arnold1998}.

Besides the mathematical benefit, there is a biological motivation in
modeling mRNA transcription by a master equation and protein synthesis
by a random differential equation, thus supposing that the transcription
product should be treated as a discrete random variable (number of mRNA
molecules) while the translation product should be treated as a continuous
random variable (density of protein molecules).
The reason behind this distinction is the large gap, typically of several
orders of magnitude, between mRNA numbers and protein numbers in the cell.
The hybrid model we propose here attempts to incorporate the discrepancy
between mRNA and protein molecules (which concerns not only their typical
numbers but also their typical lifetimes) from the very beginning, instead
of assuming that it can be ignored.
That is why, in conformity with procedures already adopted implicitly
in some of the literature but rarely spelled out (one exception is~%
\cite{Friedman2006}), we suggest to model protein number by a
continuous probability density rather than a discrete probability
distribution. 

Admittedly, this amounts to a change of paradigm, but as will be shown
here, the resulting simplifications are so substantial that they allow
us to solve the resulting model without constraints on the values of
the parameters.
Furthermore, this approach allows us to evaluate probability densities
even for very high protein numbers, with no extra effort.

\section{Model for Transcription and Translation}
\indent

Let us describe our model in more detail.
Gene transcription is described by a pair of master equations, corresponding
to two states $\{1,2\}$ of promoter activity, for a birth and death process
coupled by a telegraph-like process encoding the switch between promoter
states (generalization to a higher number of promoter states will be left
to future work):
\begin{equation}\label{eq:ME}
\begin{split}
 \frac{d\phi^{1}_{n}}{d t}~=~
 & k_{1}[\phi^{1}_{n-1}-\phi^{1}_{n}] + \rho[(n+1)\phi^{1}_{n+1} - n\phi^{1}_{n}]
   - h\phi^{1}_{n} + f\phi^{2}_{n} \,,\\ 
\frac{d\phi^{2}_{n}}{d t}~=~
 & k_{2}[\phi^{2}_{n-1}-\phi^{2}_{n}] + \rho[(n+1)\phi^{2}_{n+1} - n\phi^{2}_{n}]
   + h\phi^{1}_{n} - f\phi^{2}_{n} \,.
\end{split}
\end{equation}
The discrete random variable $n$ stands for the number of mRNA molecules
in the cell and $\phi^{j}_{n}(t)$ is the probability for finding the gene
in state number $j$ ($j=1$ or $2$) with $n$ mRNA molecules in the cell,
at time $t$; the resulting total probability will be denoted by 
$\phi_n(t)=\phi^1_n(t)+\phi^2_n(t)$.
Production of mRNA is controlled by the rates $k_1$ and $k_2$,
while its degradation is taken into account by the rate $\rho$
which is independent of the activity level of the promoter.
The switch between the two states is controlled by the rates $h$ and $f$.
Protein synthesis/degradation is governed by an RDE of the form
\begin{equation} \label{eq:RDE}
\frac{d}{d t}m_t \, = \, - \, A m_t + B n_t \,,
\end{equation}
where $m$ is a continuous random variable representing the protein number
density in the cell, $A$ and $B$ are the protein degradation and synthesis
rates, respectively, and $n$ is as before, but now with time dependence
following a stochastic Markov jump process where $n_{t+\Delta t} = n_t \pm 1$
with probability $(k_1+k_2) \Delta t$ for $+1$ and $\rho n_t
\Delta t$ for $-1$ (and $n_{t+\Delta t} = n_t$ with remaining probability):
this is consistent with the time evolution of the total probability
distribution $\phi_n$ that follows from Eq.~\eqref{eq:ME}.
With the assumption that $A$ and $B$ are constant our model focuses
on the effects of the stochasticity of the transcription process and
neglects the protein production/decay noise.

\section{Solutions of the Model}
\indent

A complete description of $n_t$ is achieved by obtaining the time dependent
solutions of the master equations \eqref{eq:ME}, and this is what we do in the 
following. However, before dealing with the master equations, let us first 
redefine the parameter space and introduce the biological quantities of the 
model, as in~\cite{Innocentini2007}, namely: the efficiency parameters
$N_1=k_1/\rho$ and $N_2=k_2/\rho$, the switching parameter $\epsilon
=(h+f)/\rho$ and the occupancy probabilities $p_1=f/(h+f)$ and $p_2=h/(h+f)$. 
Using the generating function technique \cite{vKampen2007} the coupled
master equations are transformed into a set of PDEs (partial differential
equations) for the functions $\phi^{1}(z,t)=\sum^{\infty}_{n=0}
\phi^{1}_{n}(t)z^{n}$ and $\phi^{2}(z,t)=\sum^{\infty}_{n=0}
\phi^{2}_{n}(t)z^{n}$:
\begin{equation} \label{eqp}
\begin{split}
 \frac{1}{\rho} \frac{\partial \phi^1}{\partial t}~=~
 & (z-1) \left[ N_1 \phi^1 - \frac{\partial \phi^1}{\partial z} \right]
   - \epsilon p_2 \, \phi^1 + \epsilon p_1 \, \phi^2 \,, \\
 \frac{1}{\rho} \frac{\partial \phi^2}{\partial t}~=~
 & (z-1) \left[ N_2 \phi^2 - \frac{\partial \phi^2}{\partial z} \right]
   + \epsilon p_2 \, \phi^1 - \epsilon p_1 \, \phi^2 \,.
\end{split}
\end{equation}
The probability distributions are obtained from the generating functions
\mbox{using}
\begin{equation} \label{dgen}
\begin{split}
 \phi^{1}_{n}(t)~=~\left. \frac{1}{n!} \frac{\partial \phi^{1}(z,t)}{\partial z}
 \right|_{z=0}\,,\\
 \phi^{2}_{n}(t)~=~\left. \frac{1}{n!} \frac{\partial \phi^{2}(z,t)}{\partial z}
 \right|_{z=0}\,, 
\end{split} 
\end{equation}
Introducing a new set of variables through the transformations 
$\mu=(z-1){\rm e}^{-\rho t}$ and $\nu=z-1$, Eq.~\eqref{eqp} assumes
the form
\begin{equation} \label{eqp2}
\begin{split}
 - \nu \frac{\partial \phi^1}{\partial \nu} + \nu N_1 \phi^1
 - \epsilon p_2 \, \phi^1 + \epsilon p_1 \, \phi^2
 &~=~0 \,, \\
 - \nu \frac{\partial \phi^2}{\partial \nu} + \nu N_2 \phi^2
 + \epsilon p_2 \, \phi^1 - \epsilon p_1 \, \phi^2
 &~=~0 \,,
\end{split}
\end{equation}
i.e., this transformation reduces the original set of PDEs to a set of
ODEs (ordinary differential equations), which have already been solved
in~\cite{Innocentini2007}; a similar transformation with the same purpose
has been used in~\cite{Ramos2011,Jayaprakash2009}.
Following~\cite{Innocentini2007}, the solutions of Eq.~\eqref{eqp2} are:  
\begin{subequations} \label{sol}
\begin{align}
 \phi^1(\mu,\nu)~
 =~ & F(\mu) \, p_1 \ee^{N_1 \nu} \kummerM(a,b+1,\eta) \label{sol1} \\
    & - G(\mu) (1-b) \eta^{-b} \ee^{N_1 \nu} \kummerM(a-b,1-b,\eta) \,,
 \notag \\[2mm]
 \phi^2(\mu,\nu)~
 =~ & F(\mu) \, p_2 \ee^{N_1 \nu} \kummerM(a+1,b+1,\eta) \label{sol2} \\
    & + G(\mu) (1-b) \eta^{-b} \ee^{N_1 \nu} \kummerM(1+a-b,1-b,\eta) \,,
 \notag
\end{align}
\end{subequations}
where $F$ and $G$ are arbitrary functions that must be determined from
the initial conditions, where we note that $t=0$ corresponds to $\nu=\mu$.
The symbol $\kummerM$ stands for the Kummer M function~\cite{AandS} with
parameters $a = \epsilon p_{2}$, $b = \epsilon$ and $\eta = (N_{2}-N_{1})\nu$.

In order to determine $F$ and $G$ we will use matrix and vector notation
to rewrite the solutions of Eq.~\eqref{sol} as $\vec{\phi} (\mu,\nu)
= U(\nu) \vec{F}(\mu)$, where $\vec{\phi} =(\phi^{1},\phi^{2})^\transp$
and $\vec{F}=(F,G)^\transp$ (where $.^\transp$ means matrix transposition);
then the entries of the matrix $U(\nu)$ are
\begin{equation}
 \begin{split}
  & U_{1,1}~=~p_{1} \, \ee^{N_{1}\nu} \kummerM(a,b+1,\eta) \,, \\[0.5ex]
  & U_{1,2}~=~- (1-b) \eta^{-b} \, \ee^{N_{1}\nu} \kummerM(a-b,1-b,\eta) \,,
  \\[0.5ex]
  & U_{2,1}~=~p_{2} \, \ee^{N_{1}\nu} \kummerM(a+1,b+1,\eta) \,, \\[0.5ex]
  & U_{2,2}~=~(1-b) \eta^{-b} \, \ee^{N_{1}\nu} \kummerM(1+a-b,1-b,\eta) \,.
 \end{split}
\end{equation}
Inverting the relation $\, \vec{\phi}(\mu,\nu) = U(\nu) \vec{F}(\mu) \,$
gives $\, \vec{F}(\mu) = U(\nu)^{-1} \vec{\phi}(\mu,\nu)$, and setting
$\nu = \mu$, we obtain an expression for $\vec{F}(\mu)$ in terms of
the initial conditions.
Thus we have to compute the inverse of the matrix $U(\nu)$, which requires
calculating its determinant.
At a first glance, it might appear difficult to find a compact formula
for that, since it involves products of Kummer functions.
Fortunately, the well known relations for Kummer functions,
especially the one concerning the Wronskian (relations 13.1.20
in~\cite{AandS}), allow us to obtain a simple expression for this
determinant:
\begin{equation}\label{det}  
 \det(U(\nu))~=~\frac{\ee^{-(N_{1}+N_{2})\nu} \eta^{\epsilon}}{1-\epsilon} \,.
\end{equation}
Putting everything together, we obtain the time dependent probability
distributions that solve Eq.~\eqref{eq:ME} and will serve as input to
solve Eq.~\eqref{eq:RDE}.

Considering any given perturbation $n_t$ as input, the ODE~\eqref{eq:RDE}
governing the protein dynamics is easily solved by applying the standard
integral formula from the theory of ODEs. Introducing the dimensionless
parameters $\tau = \rho\,t$, $\alpha = A/\rho$ and $\beta = B/\rho$,
the solution reads
\begin{equation} \label{eq:RDESOLUTION}
 m_\tau~=~m_0 \, \ee^{-\alpha \tau} + \beta \, \ee^{-\alpha\tau} 
 \int_{0}^{\tau} n_{\tau'} \, \ee^{\alpha\tau'} \dd \tau' \,,
\end{equation}
where the integral is an ordinary Riemann integral (applied to the
product of a step function by an exponential function) and $m_0=m(0)$.
In the present case, where both $n_\tau$ and $m_\tau$ are stochastic
processes, we can interpret this formula as an operator that maps the
process $n_\tau$ (for mRNA number) to the process $m_\tau$
(for protein number density), sample by sample.

Recalling that the ultimate goal is to compute the probability density
of the protein population, say $\mathscr{P}(\tau,m)$, the traditional
method consists in randomly generating stochastic processes $n_\tau$
for mRNA number, applying the previous integral formula to produce
corresponding stochastic processes $m_\tau$ for protein number density
and looking at the resulting statistics.
Here, and this is perhaps the central point of the present paper, we
propose a different procedure: since the solution of Eq.~\eqref{eq:ME}
has already provided us with a probability distribution for mRNA number,
it suffices to take its push-forward, in the sense of measure theory,
under the operator defined by solving Eq.~\eqref{eq:RDESOLUTION} to
directly obtain the corresponding probability distribution for protein
number density, without having to resort to random process generation.
To describe how to compute the push-forward, let us consider the
integral on the rhs of Eq.~\eqref{eq:RDESOLUTION}. Dividing the
interval $[0,\tau]$ in $p$ subintervals we have:
\begin{equation} \label{eq:RDESOLPART}
 \int_{0}^{\tau} n_{\tau'} \, \ee^{\alpha\tau'} \dd \tau'~
 =~\sum_{q=0}^{p-1} \int_{\tau_q}^{\tau_{q+1}}
   n_{\tau'} \, \ee^{\alpha\tau'} \dd \tau',
\end{equation}
where $\tau_{0}=0 $ and $\tau_{p}=\tau$. If the partition is
sufficiently fine (i.e., for $p$ sufficiently large), the function $n_\tau$ will
be constant on each subinterval and the integral can be performed explicitly:
\begin{equation} \label{eq:RDESOLAPPROX}
 m_\tau~=~m_0 \, \ee^{-\alpha \tau} + \frac{\beta}{\alpha} \ee^{-\alpha\tau}
          \sum_{q=0}^{p-1} n_{\tau_{q}}
          (\ee^{\alpha\tau_{q+1}}-\ee^{\alpha\tau_{q}}).
\end{equation}
Otherwise, i.e., for smaller values of~$p$, Eq.~\eqref{eq:RDESOLAPPROX}
provides only a ``rectangular'' or ``piecewise constant'' approximation of the
integral in Eq.~\eqref{eq:RDESOLPART} since it amounts to replacing, on each
of the subintervals $[\tau_{q},\tau_{q+1}]$, the step function $n_{\tau'}$
by a constant (here chosen to be its value at the left endpoint):
\begin{equation}
 \int_{\tau_{q}}^{\tau_{q+1}} n_{\tau'} \, \ee^{\alpha\tau'} \dd \tau'
 \approx n_{\tau_{q}} \int_{\tau_{q}}^{\tau_{q+1}} \ee^{\alpha\tau'} \dd \tau'~
 =~\frac{n_{\tau_{q}}}{\alpha} \, \bigg|_{\tau_q}^{\tau_{q+1}}.
\end{equation}
Of course, a ``trapezoidal'' or ``piecewise linear'' approximation is more
precise: it consists in replacing this expression by
\begin{equation}
\int_{\tau_{q}}^{\tau_{q+1}} n_{\tau'} \, \ee^{\alpha\tau'} \dd \tau'
 \approx \int_{\tau_{q}}^{\tau_{q+1}} (a \tau' + b) \,
         \ee^{\alpha\tau'} \dd \tau'
 = \frac{1}{\alpha} \, ( a \tau' + b - \frac{a}{\alpha}) \,
   \ee^{\alpha\tau'} \, \bigg|_{\tau_q}^{\tau_{q+1}},
\end{equation}
where $a$ and $b$ are determined by solving the equations $\, n_{\tau_{q}}
= a \tau_q + b \,$ and $\, n_{\tau_{q+1}} = a \tau_{q+1} + b$.

In order to obtain a sample path for the process $m_\tau$ using these
formulas, it suffices to represent a sample path for the process $n_\tau$
by the ``shrunk'' numerical sequence $(n_0,\ldots,n_{p-1})$, the only
modification being that we must now allow consecutive numbers to differ
by more than $\pm 1$.
Finally, to make our sample space finite, we also introduce a cutoff~$L$
and impose that all $n_q$ should be $\leqslant L$.
For instance, by choosing $L$ so large that the probability of $n_q> L$ is 
smaller than $10^{-20}$, say, we can certainly neglect all values higher 
than $L$ and restrict the set of possible values for $n_q$ to the finite set 
$\{0,1,\ldots,L-1,L\}$; then the space of sequences has $(L+1)^p$ elements.

Now, Eq.~\eqref{eq:RDESOLAPPROX} provides a map from this space
of sequences $(n_{0},\ldots,n_{p-1})$ to that of numbers $m_\tau$.
Using this mapping we define the push-forward probability on the set
of possible values of $m_\tau$ by

\begin{equation} \label{probproduct}
 \Prob \bigl( m_\tau = m_\tau(n_{0},\ldots,n_{p-1}) \bigr)
 = \Phi(n_0;\ldots;n_{p-1}),
 \end{equation}
where
\begin{equation}
 \Phi(n_0;\ldots;n_{p-1})
 = \Phi^1(n_0;\ldots;n_{p-1}) + \Phi^2(n_0;\ldots;n_{p-1})
\end{equation}
is the total joint probability distribution for finding $n_q$ mRNA molecules
at times $\tau_q$ $(q=0,\ldots,p-1)$, whereas $\Phi^1(n_0;\ldots;n_{p-1})$
and $\Phi^2(n_0;\ldots;n_{p-1})$ encode the joint probability distributions
for finding $n_q$ mRNA molecules at times $\tau_q$ \linebreak $(q=0,...,p-1)$
with the gene in promoter state $1$ and $2$, respectively.
In general, \linebreak such joint probability distributions are difficult to
obtain, but in our case, the mRNA process governed by the master equations~%
\eqref{eq:ME} is markovian and therefore we can compute the joint probabilities
in terms of conditional probabilities, according to the iterated Chapman-%
Kolmogorov equation:
\begin{equation} \label{pcond1}
\begin{split}
 \Phi^1(n_0;\ldots;n_{p-1}) = \sum_{j_0,\ldots,j_{p-2}=1}^{2}
  & \Phi(n_{p-1},\tau_{p-1},1|n_{p-2},\tau_{p-2},j_{p-2}) \ldots \\[-1ex]
  & \ldots \Phi(n_{1},\tau_{1},j_1|n_{0},\tau_{0},j_0) \,
           \phi_{n_0}^{j_0}(\tau_0) \,, \\
 \Phi^2(n_0;\ldots;n_{p-1}) = \sum_{j_0,\ldots,j_{p-2}=1}^{2}
  & \Phi(n_{p-1},\tau_{p-1},2|n_{p-2},\tau_{p-2},j_{p-2}) \ldots \\[-1ex]
  & \ldots \Phi(n_{1},\tau_{1},j_1|n_{0},\tau_{0},j_0) \,
    \phi_{n_0}^{j_0}(\tau_0) \,,
\end{split}
\end{equation}
where, as before, $\phi_{n_0}^{j_0}(\tau_0)$ is the probability to find the
gene in the state $j_0$ and with $n_0$ mRNA molecules in the cell, at time
$\tau_0$.
The quantity $\Phi(n_{q'},\tau_{q'},j'|n_q,\tau_q,j)$ is the conditional
probability of finding $n_{q'}$ mRNA molecules at time $\tau_{q'}$ and with
the gene in state $j'$ provided there were $n_q$ mRNA molecules at time
$\tau_q$ and with the gene in state $j$, where $\tau_q < \tau_{q'}$,
$q,q'=1,\ldots,p-1$ and $j,j'=1,2$.
These conditional probabilities can be obtained from the solutions
of the master equations~\eqref{sol}.
To this end, one has to take as initial condition the generating
function encoding the information that, at time $\tau_q$, the system
has exactly $n_q$ particles  and with probability $1$ is in one of
the two promoter states, say 1 or~2. \linebreak
Such a generating function has one component equal to $0$ whereas the
other is given by $(1+\mu)^{n_q}$, i.e.,
\begin{equation} \label{ini1a}
 (\phi^1(\mu,\nu),\phi^2(\mu,\nu)) = ([1+\mu]^q,0)
 \quad \mbox{with} \quad \nu=\mu
\end{equation}
for promoter in state~$1$ and
\begin{equation} \label{ini1b}
 (\phi^1(\mu,\nu),\phi^2(\mu,\nu)) = (0,[1+\mu]^q)
 \quad \mbox{with} \quad \nu=\mu
\end{equation}
for promoter in state~$2$. In the $(z,\tau)$ variables, the non-vanishing
component takes the form
\begin{equation} \label{ini2}
 \left( 1 + (z-1) \ee^{-(\tau-\tau_q)} \right)^{n_q}
\end{equation}  
since here the initial time is $\tau_q$, rather than $0$. 

Regarding the validity of Eq.~\eqref{probproduct}, it is important to note
that according to the general definition of the push-forward of probabilities,
one should really take the sum of the probabilities corresponding to all
sequences $(n_0,\ldots,n_{p-1})$ producing the same value of $m_\tau$.
However, Eq.~\eqref{eq:RDESOLAPPROX} implies that, generically, any
two different sequences will give different values (more precisely, this will
be the case if the intermediate times $\tau_1,\ldots,\tau_{p-1}$ are chosen
such that the differences of exponentials $\, \ee^{\alpha\tau_{q+1}} -
\ee^{\alpha\tau_{q}}$, $q=0,\ldots,p-1$, are linearly independent
over the integers).

For the sake of greater clarity, and to illustrate how the conditional
probabilities are obtained from the explicit solution~\eqref{sol} of
the master equations with the appropriate initial conditions (see Eqs~%
\eqref{ini1a},\eqref{ini1b} and \eqref{ini2} above), let us consider the
simplest example: $p=2$ and $L=1$.
Here, the sample space has four elements, namely, $(0,0)$, $(0,1)$, 
$(1,0)$ and $(1,1)$, and in general each of these sequences will
produce a different number $m_\tau$. Therefore, the probability assigned
to each of these values $m_\tau$ is equal to the joint probability
assigned to the corresponding sequence $(n_0,n_1)$, summed
over the two possible promoter states,
\begin{equation}
\Prob(m_\tau=m_\tau(n_0,n_1))= \Phi^{1}(n_0;n_{1})+\Phi^{2}(n_0;n_{1}).
\end{equation}
Specializing Eq.~\eqref{pcond1} to the case $p=2$, we see that
these joint probabilities are
\begin{equation}
 \begin{array}{c}
  \Phi^1(n_0;n_1) = \Phi(n_1,\tau_1,1|n_0,\tau_0,1) \, \phi^1_{n_0}(\tau_0)
                  + \Phi(n_1,\tau_1,1|n_0,\tau_0,2) \,
                    \phi^2_{n_0}(\tau_0) \,, \\[1ex]
  \Phi^2(n_0;n_1) = \Phi(n_1,\tau_1,2|n_0,\tau_0,1) \, \phi^1_{n_0}(\tau_0)
                  + \Phi(n_1,\tau_1,2|n_0,\tau_0,2) \,
                    \phi^2_{n_0}(\tau_0) \,,
 \end{array}
\end{equation}
where, as before, the conditional probabilities $\Phi(n_1,\tau_1,j_1|%
n_0,\tau_0,j_0)$ take into account the promoter states.
To exemplify how these are obtained from the solutions of the master
equations, let us, by way of example, focus on the conditional
probability \mbox{$\Phi(n_1=5,\tau_1,j_1=1|n_0=10,\tau_0,j_0=1)$.}
This means that we are considering the situation where, at time
$\tau_0$, the system has 10 mRNA molecules and the promoter
is found in the state 1, corresponding to the initial
condition
\begin{equation} \label{ini3}
 (\phi^1(\mu,\nu),\phi^2(\mu,\nu)) = ([1+\mu]^{10},0)
 \quad \mbox{with} \quad \nu=\mu,
\end{equation}
or in the $(z,\tau)$ variables,
\begin{equation} \label{ini4}
 (\phi^1(z,\tau_1),\phi^2(z,\tau_1))
 =  \left( \left[1+(z-1) \, \ee^{-(\tau_1-\tau_0)} \right]^{10},0 \right).
\end{equation}
Using Eq.~\eqref{ini3} to determine the vector
\begin{equation} \label{ini5}
 \vec{F}(\mu) = U(\nu)^{-1}\vec{\phi}(\mu,\nu) \Bigr|_{\nu=\mu},
\end{equation}
substituting the entries of this vector in Eq.~\eqref{sol1} for $\phi^1$ and
returning to the variables $(z,\tau)$, we arrive at the generating
function, let's say $\Psi(z,\tau)$, of the conditional probabilities
\mbox{$\Phi(n_1,\tau_1,j_1=1|n_0=10,\tau_0,j_0=1)$}, from which the
conditional probability under consideration can be obtained by taking
derivatives, as follows:
\begin{equation}
 \Phi(n_1=5,\tau_1,j_1=1|n_0=10,\tau_0,j_0=1)
 = \frac{1}{5!} \frac{\partial^5 \Psi(z,\tau)}{\partial z^5} \Bigr|_{z=0}.
\end{equation}
When the system at initial time $\tau_0$ is in promoter state 2
rather than 1, we have to switch the two components in the vector of Eqs~%
\eqref{ini3} and~\eqref{ini4}, use the entries of this vector to determine
$\vec{F}$, according to Eq.~\eqref{ini5}, and again apply
Eq.~\eqref{sol1} for~$\phi^1$ to obtain the generating function for the
conditional probability \linebreak $\Phi(n_1=5,\tau_1,j_1=1|%
n_0=10,\tau_0,j_0=2)$.
And finally, to compute the conditional probabilities
$\Phi(n_1=5,\tau_1,j_1=2|n_0=10,\tau_0,j_0)$, with $j_0 = 1 \,
\mbox{or} \, 2$, we proceed in the same way, the only difference
being that instead of using Eq.~\eqref{sol1} for $\phi^1$ we use
Eq.~\eqref{sol2} for $\phi^2$.

From Eq.~\eqref{probproduct} the probability density for protein number
is obtained as the limit
\begin{equation} \label{pdens}
 \mathscr{P}(\tau,m)=\lim_{L,p\to\infty}
 \Prob(m_\tau(n_{0},\ldots,n_{p-1})),
\end{equation}
where $\tau_{q+1}-\tau_{q} \to 0$ as $p \to \infty$ in such a way that the
product $p\,(\tau_{q+1}-\tau_{q})$ remains finite.
The computational implementation of this limit is obtained by approximating
the probability density by a histogram.

Finally, to consider arbitrarily long times, we take advantage of the
fact that Eq.~\eqref{eq:RDE} is autonomous and hence its solutions have
a composition property, namely:
\begin{equation}
 m_{\tau,\tau}= \id \quad , \qquad 
 m_{\tau,\tau'}\circ m_{\tau',\tau''}=m_{\tau,\tau''} \,.
\end{equation}
These formulas are obtained from the general solution of the initial
value problem with $m(\tau')=m_{\tau'}$ ($\tau'<\tau$),
\begin{equation} \label{eq:SOLUTMAP}
 m_{\tau,\tau'} = m_{\tau'} \, \ee^{-\alpha(\tau-\tau')}
 +\beta \int_{\tau'}^{\tau} n_{\tau''} \, \ee^{-\alpha(\tau-\tau'')} \dd\tau'',
\end{equation}
which defines a family of transformations acting on the set of initial 
conditions.
By iteration, it follows that the solution may be written as
$m_\tau = m_{\tau_p,\tau_{p-1}} \circ\cdots\circ m_{\tau_1,\tau_0}$,
where $\{\tau_0=0,\ldots,\tau_p=\tau\}$ is any subdivision of
the time interval $[0,\tau]$ and each $m_{\tau_{q+1},\tau_q}$ is
given by~Eq.~\eqref{eq:SOLUTMAP}, with the initial condition
$m(\tau_q)=m_{\tau_q}$ having probability density
$\mathscr{P}(\tau_q,m)$, for $q=0,\ldots,p-1$.

\section{Moments of mRNA number and protein number distribution}

The time dependent mRNA moments can be obtained directly from the
solutions of Eqs~\eqref{eqp2} given in Eqs~\eqref{sol}, by transforming
back to the original $(z,\tau)$ variables and taking derivatives of these
generating functions with respect to the variable $z$ at $z=1$:
\begin{equation}
 \langle n^{(r)}_{\tau} \rangle_j^{}
 = \left( z \frac{\partial}{\partial z} \right)^{\!r} \phi^j(z,\tau)
   \Big|_{z=1} \quad (j=1,2).
\end{equation}
Alternatively, we can view each of these moments as the solution of
its own system of ordinary differential equations, obtained by applying the
operator $(z \, \partial/\partial z)^r |_{z=1}$ directly to the system
of partial differential equations~\eqref{eqp}, rather than its solutions.
This is the procedure we shall adopt in what follows, for the first
two moments.

As a preliminary step, we note that taking $r=0$ (which amounts to
simply evaluating Eq.~\eqref{eqp} at $z=1$) gives, for the promoter
state occupancy probabilities
\begin{equation}
 \pi_j(\tau) = \sum_{n \geq 0} \phi_n^j(\tau) = \phi^j(\tau,z=1)
 \quad (j=1,2),
\end{equation}
the following system of differential equations,
\begin{equation} \label{sysocpr}
 \begin{array}{c}
  {\displaystyle
   \D{\pi_1} = - \, \epsilon p_2 \, \pi_1 + \epsilon p_1 \, \pi_2} \,, \\[2ex]
  {\displaystyle
   \D{\pi_2} = \epsilon p_2 \, \pi_1 - \epsilon p_1 \, \pi_2} \,.
 \end{array}
\end{equation}
Its solution is immediate,
\begin{equation} \label{solsysocpr}
 \begin{array}{c}
  \pi_1(\tau) = p_1 + (\pi_1(0) - p_1) \, \ee^{-\epsilon \tau}, \\[1ex]
  \pi_2(\tau) = p_2 + (\pi_2(0) - p_2) \, \ee^{-\epsilon \tau},
 \end{array}
\end{equation}
provided we take into account that $p_1+p_2 = 1$: this will imply that
the constraint $\, \pi_1(\tau) + \pi_2(\tau) = 1 \,$ is conserved (it holds
for all $\tau$ provided it holds for the initial condition, i.e., for $\tau=0$)
and allow us to interpret the coefficients $p_j$ as the asymptotic promoter
state occupancy probabilities:
\begin{equation}
 p_j = \lim_{\tau \to \infty} \pi_j(\tau) \quad (j=1,2).
\end{equation}

\subsection{Mean values}
\indent

Considering the case $r=1$, we apply the operator $(z \, \partial/\partial z)$
to Eqs~\eqref{eqp} and evaluate at $z=1$ to obtain, for the mean partial mRNA
numbers
\begin{equation}
 \langle n_{\tau}^{(1)} \rangle_j^{} = \sum_{n \geq 0} n \, \phi_n^j(\tau)
 = \Bigl( z \frac{\partial}{\partial z} \Bigr) \phi^j(z,\tau) \Big|_{z=1}
 \quad (j=1,2),
\end{equation}
the following system of differential equations,
\begin{equation} \label{sysmean}
\begin{array}{c}
 \dfrac{\dd}{\dd\tau} \langle n^{(1)}_\tau \rangle_1^{}
 = - (1+\epsilon p_2) \langle n^{(1)}_\tau \rangle_1^{}
   + \epsilon p_1 \langle n^{(1)}_\tau \rangle_2^{} + N_1 \pi_1(\tau) \,,
 \\[2ex]
 \dfrac{\dd}{\dd\tau} \langle n^{(1)}_\tau \rangle_2^{}
 = - (1+\epsilon p_1) \langle n^{(1)}_\tau \rangle_2^{}
   + \epsilon p_2 \langle n^{(1)}_\tau \rangle_1^{} + N_2 \pi_2(\tau) \,.
\end{array} 
\end{equation}
The corresponding differential equation for the mean total mRNA number
\begin{equation}
 \langle n_{\tau}^{(1)} \rangle
 = \langle n_{\tau}^{(1)} \rangle_1^{} + \langle n_{\tau}^{(1)} \rangle_2^{}
\end{equation}
is obtained by summing over $j$:
\begin{equation} \label{sysmeantot}
 \dfrac{\dd}{\dd\tau} \langle n_{\tau}^{(1)} \rangle
 = - \langle n_{\tau}^{(1)} \rangle  + N_1 \pi_1(\tau) + N_2 \pi_2(\tau) \,.
\end{equation}
Note that we can solve this equation without having to solve the full
system~\eqref{sysmean}.
Namely, introducing the constants
\begin{equation}
 \bar{N} = N_1 p_1 + N_2 p_2 \quad , \quad
 \Delta N = N_1 - N_2 \,,
\end{equation}
we get from Eq.~\eqref{solsysocpr}
\[
 N_1 \pi_1(\tau) + N_2 \pi_2(\tau)
 = \bar{N} + \Delta N \, (\pi_1(0) - p_1) \, \ee^{-\epsilon \tau},
\]
and this can be used to integrate Eq.~\eqref{sysmeantot}, after putting it
in the form
\[
 \ee^{-\tau} \D{\bigl( \ee^\tau \langle n^{(1)}_\tau \rangle \bigr)}
 = N_1 \pi_1(\tau) + N_2 \pi_2(\tau) \,.
\]
The solution is
\begin{equation} \label{ncopy} 
 \langle n^{(1)}_\tau \rangle
 = \bar{N} + (\langle n^{(1)}_0 \rangle - \bar{N}) \, \ee^{-\tau}
   + \frac{\Delta N}{1-\epsilon} \bigl( \pi_1(0) - p_1 \bigr)
     \bigl( \ee^{-\epsilon \tau} - \ee^{-\tau} \bigr) \,,
\end{equation}
with the asymptotic value
\begin{equation}
 \langle n^{(1)}_\infty \rangle
 = \lim_{\tau \to \infty} \langle n^{(1)}_\tau \rangle = \bar{N} \,.
\end{equation}
For later use, we record here the complete solution of the system~%
\eqref{sysmean} because it will be needed at the next stage; it reads
\begin{equation} \label{n1mrna}
 \begin{split}
 \langle n^{(1)}_\tau \rangle_1^{} = \;
 & \langle n^{(1)}_\infty \rangle_1^{}
   + \frac{\epsilon \, \Delta N \, p_1 (\pi_1(0) - p_1)}{1-\epsilon} \,
     \ee^{-\tau} \\
 & + \frac{[\epsilon(\Delta N \, p_1 - N_1) + N_1](\pi_1(0) - p_1)}%
          {1-\epsilon} \, \ee^{-\epsilon\tau} \\
 & - \frac{\epsilon \, \Delta N (\pi_1(0) - p_1)(\pi_2(0) - p_1)}%
          {1+\epsilon} \, \ee^{-(1+\epsilon)\tau}
 \end{split}
\end{equation}
for the partial mean value when the gene is in the state 1, and
\begin{equation} \label{n2mrna}
\begin{split}
 \langle n^{(1)}_\tau \rangle_2^{} = \;
 & \langle n^{(1)}_\infty \rangle_2^{}
   + \frac{\epsilon \, \Delta N \, p_2 (\pi_1(0)\!-\!p_1)}{1-\epsilon} \,
     \ee^{-\tau} \\
 & - \frac{[\epsilon(\Delta N \, p_1 - N_1) + N_2](\pi_1(0) - p_1)}%
          {1-\epsilon} \, \ee^{-\epsilon\tau} \\
 & + \frac{\epsilon \, \Delta N (\pi_1(0) - p_1)(\pi_2(0) - p_1)}%
          {1+\epsilon} \, \ee^{-(1+\epsilon)\tau}
\end{split}
\end{equation}
for the partial mean value when the gene is in the state 2, with the
asymptotic values
\begin{equation} \label{solmeanpar}
 \langle n^{(1)}_\infty \rangle_1^{}
 = N_1 p_1 + \frac{\epsilon N_2 p_2}{1+\epsilon}
 \quad , \quad
 \langle n^{(1)}_\infty \rangle_2^{}
 = N_2 p_2 + \frac{\epsilon N_1 p_1}{1+\epsilon} \,.
\end{equation}

The ordinary differential equation governing the mean protein number density
is obtained by averaging Eq.~\eqref{eq:RDE} which, in terms of the rescaled
variables, gives
\begin{equation} \label{eq:MEANEQ}
 \dfrac{\dd}{\dd \tau} \langle m_\tau \rangle
 = - \, \alpha \, \langle m_\tau \rangle
   + \beta \, \langle n^{(1)}_\tau \rangle. 
\end{equation}
The solution of this equation is as in Eq.~\eqref{eq:RDESOLUTION}:
\begin{equation} \label{eq:mean}
 \langle m_\tau \rangle = \langle m_0 \rangle \, \ee^{-\alpha\tau}
 + \beta \, \ee^{-\alpha\tau} \int_{0}^{\tau} \langle n^{(1)}_{\tau'} \rangle \,
   \ee^{\alpha\tau'} \dd \tau'. 
\end{equation} 
Using Eqs~\eqref{ncopy} and~\eqref{eq:mean}, we integrate this to find
\begin{equation} \label{mcopy}
\begin{split}
 \langle m_\tau \rangle~=
 &~\langle m_0 \rangle \, \ee^{-\alpha\tau}
   + \bar{N} \frac{\beta}{\alpha} (1 - \ee^{-\alpha\tau})
   + \beta (\langle n^{(1)}_0 \rangle - \bar{N})
     \left(\frac{\ee^{-\tau} - \ee^{-\alpha\tau}}{\alpha-1} \right) \\
 &~+ \Delta N \frac{\beta}{1-\epsilon} (\pi_1(0) - p_1)
     \left( \frac{\ee^{-\epsilon \tau} - \ee^{-\alpha\tau}}{\alpha-\epsilon}
   - \frac{\ee^{-\tau} - \ee^{-\alpha\tau}}{\alpha-1} \right).
\end{split}
\end{equation}
with the asymptotic value
\begin{equation}
 \langle m_\infty \rangle  = \lim_{\tau \to \infty} \langle m_\tau \rangle
 = \bar N \frac{\beta}{\alpha} \,.
\end{equation}

\subsection{Variance}
\indent

Passing to the case $r=2$, we apply the operator $(z \, \partial/\partial z)$
to Eqs~\eqref{eqp} twice and evaluate at $z=1$ to obtain, for the partial
second moments
\begin{equation}
 \langle n_{\tau}^{(2)} \rangle_j^{} = \sum_{n \geq 0} n^2 \, \phi_n^j(\tau)
 = \Bigl( z \frac{\partial}{\partial z} \Bigr)^{\!2} \phi^j(z,\tau) \Big|_{z=1}
 \quad (j=1,2),
\end{equation}
the following system of ordinary differential equations,
\begin{equation} \label{syssecm}
\begin{array}{c}
 \dfrac{\dd}{\dd\tau} \langle n^{(2)}_\tau \rangle_1^{}
 = - \, 2 \langle n^{(2)}_\tau \rangle_1^{}
   + (2 N_1 + 1 - \epsilon p_2) \langle n^{(1)}_\tau \rangle_1^{}
   + \epsilon p_1 \langle n^{(1)}_\tau \rangle_2^{} + N_1 \pi_1(\tau), \\[2ex]
 \dfrac{\dd}{\dd\tau} \langle n^{(2)}_\tau \rangle_2^{}
 = - \, 2 \langle n^{(2)}_\tau \rangle_2^{}
   + (2 N_2 + 1 - \epsilon p_1) \langle n^{(1)}_\tau \rangle_2^{}
   + \epsilon p_2 \langle n^{(1)}_\tau \rangle_1^{} + N_2 \pi_2(\tau).
\end{array} 
\end{equation}
The corresponding differential equation for the total second moment
\begin{equation}
 \langle n_{\tau}^{(2)} \rangle
 = \langle n_{\tau}^{(2)} \rangle_1^{} + \langle n_{\tau}^{(2)} \rangle_2^{}
\end{equation}
is obtained by summing over $j$:
\begin{equation} \label{syssecmtot}
 \dfrac{\dd}{\dd\tau} \langle n_{\tau}^{(2)} \rangle
 = - \, 2 \langle n^{(2)}_\tau \rangle
   + (2 N_1 + 1) \langle n^{(1)}_\tau \rangle_1
   + (2 N_2 + 1) \langle n^{(1)}_\tau \rangle_2
   + N_1 \pi_1(\tau) + N_2 \pi_2(\tau).
\end{equation}
Equivalently, we can derive a differential equation directly for the variance
\begin{equation}
 V(n_\tau) = \langle n^{(2)}_\tau \rangle - \langle n^{(1)}_\tau \rangle^2
\end{equation}
by using Eq.~\eqref{sysmeantot} to deduce that
\[
 \dfrac{\dd}{\dd\tau} \langle n^{(1)}_\tau \rangle^2
 = 2 \langle n^{(1)}_\tau \rangle \,
   \dfrac{\dd}{\dd\tau} \langle n^{(1)}_\tau \rangle
 = - 2 \langle n^{(1)}_\tau \rangle^2
   + 2 \langle n^{(1)}_\tau \rangle (N_1 \pi_1(\tau) + N_2 \pi_2(\tau)) 
\]
and subtracting this result from Eq.~\eqref{syssecmtot} to arrive at
\begin{equation} \label{varmrna2}
\begin{split}
 \dfrac{\dd}{\dd\tau}V(n_\tau) = \;
 & - 2 V(n_\tau) + \langle n^{(1)}_\tau \rangle
     [1 - 2 (N_1 \pi_1(\tau) + N_2 \pi_2(\tau))] \\
 & + 2 N_1 \langle n^{(1)}_\tau \rangle_1^{}
   + 2 N_2 \langle n^{(1)}_\tau \rangle_2^{}
   + N_1 \pi_1(\tau) + N_2 \pi_2(\tau).
\end{split} 
\end{equation}
Again, we can solve Eqs~\eqref{syssecmtot} and~\eqref{varmrna2} without
having to solve the full system~\eqref{syssecm}, but here we now need the
full solution of the system~\eqref{sysmean}, Eqs~\eqref{n1mrna} and~%
\eqref{n2mrna}.
For the variance, this solution has the following structure:
\begin{equation}\label{vns}
 V(n_{\tau})
 = A_1 + B_1 \ee^{-\tau} + C_1 \ee^{-2\tau}
   + D_1 \ee^{-\epsilon\tau} +  E_1 \ee^{-(1+\epsilon)\tau}
   + F_1 \ee^{-2\epsilon\tau},
\end{equation}
with coefficients given by:
\begin{equation}
 \begin{array}{c}
 {\displaystyle A_1 =
  \bar{N} + \frac{(\Delta N)^2 \, p_1(1-p_1)}{1+\epsilon} \,,} \\[3ex]
 {\displaystyle B_1 =
  - \, \frac{\epsilon \, \Delta N \, (\pi_1(0)-p_1)}{1-\epsilon} \,,} \\[3ex]
 {\displaystyle C_1 =
  - \, \frac{\epsilon \, (\Delta N)^2 \, (\pi_1(0)-p_1) \,
             [2\pi_1(0) - \epsilon (\pi_1(0)-p_2) - 1]}%
            {(1-\epsilon)^2(2-\epsilon)} \,,} \\[3ex]
 {\displaystyle D_1 =
  \Delta N \, (\pi_1(0)-p_1)
  \left[\frac{1}{1-\epsilon} + \frac{2 \, \Delta N \, (1-2p_1)}%
                                    {2-\epsilon} \right] \,,} \\[3ex]
 {\displaystyle E_1 =
 \frac{2 \, \epsilon \, (\Delta N)^2 \, (\pi_1(0)-p_1) \,
       [2\pi_1(0) - \epsilon (1-2p_1) - 1]}%
      {(1+\epsilon)(1-\epsilon)^2} \,,} \\[3ex]
 {\displaystyle F_1 =
 - \, \frac{(\Delta N)^2 \, (\pi_1(0)-p_1)^2}{(1-\epsilon)^2} \,.}
 \end{array}
\end{equation}

Our final goal will be to analyze the variance of the protein number density,
\begin{equation}
 V(m_\tau) = \langle m^2_\tau \rangle - \langle m_\tau \rangle^2 \,.
\end{equation}
Using the solution for
$\langle m_\tau \rangle$ in its integral representation, Eq.~\eqref{eq:mean},
the expression for $\langle m_\tau \rangle^2$ is
\begin{equation} \label{mcopy2}
\begin{split}
 \langle m_\tau \rangle^2 = \;
 & \langle m_0 \rangle^2 \, \ee^{-2\alpha\tau} + 2\beta\,\ee^{-2\alpha\tau}
   \int_0^\tau \langle m_0 \rangle \langle n^{(1)}_{\tau'} \rangle \,
   \ee^{\alpha\tau'} d\tau' \\
 & + \beta^2 \ee^{-2\alpha\tau} \int_0^\tau \int_0^\tau
     \langle n^{(1)}_{\tau'} \rangle \langle n^{(1)}_{\tau''} \rangle \,
     \ee^{\alpha(\tau'+\tau'')} d\tau' d\tau''.
\end{split}
\end{equation}
The expression for $\langle m^2_\tau \rangle$ is obtained by first squaring
Eq.~\eqref{eq:RDESOLUTION} and then averaging, leading to:
\begin{equation} \label{secp}
\begin{split}
 \langle m^2_\tau \rangle = \;
 & \langle m^2_0 \rangle \, \ee^{-2\alpha\tau} + 2\beta\,\ee^{-\alpha\tau}
   \int_0^\tau \langle m_0 \, n_{\tau'} \rangle \, \ee^{\alpha\tau'} d\tau' \\
 & +\beta^2\ee^{-2\alpha\tau} \int_0^\tau \int_0^\tau \langle n_{\tau'}
   n_{\tau''} \rangle \, \ee^{\alpha(\tau'+\tau'')} d\tau' d\tau''.
\end{split}
\end{equation}
With these expressions at hand and in view of the fact that $\langle m_0
n_\tau \rangle = \langle m_0 \rangle \langle n^{(1)}_\tau \rangle$, which
means that the initial condition $m_0$ for protein number is independent
of the mRNA process $n_\tau$, we arrive at an explicit expression for the
variance of protein number:
\begin{equation}\label{mvar}
 V(m_\tau) = \ee^{-2\alpha\tau} \left[ V(m_0) + \beta^2
 \underbrace{\int_0^\tau \int_0^\tau \ee^{\alpha(s+s')}
 \bigl( \langle n_{s} n_{s'} \rangle - \langle n^{(1)}_s \rangle
        \langle n^{(1)}_{s'} \rangle) \, ds \, ds'}_{I_\tau} \right],
\end{equation}
where $\langle n_s n_{s'} \rangle$ is the mRNA correlation function.
Using the tower property of the conditional expectation and the Markov
property of the solution of the master equation, we get, for $s>s'$
\begin{equation}\label{nsns}
\begin{split}
 \langle n_s n_{s'} \rangle
 &= \sum_{n'} \sum_{n} \sum_{j} n \, n' \, \Phi^{j}(n',s';n,s) \\
 &= \sum_{n',j'} n' \left[ \sum_{n,j} n \,\Phi(n,s,j|n',s',j')\right]
    \, \phi_{n'}^{j'}(s') \\
 &= \sum_{n',j'} n' \, \langle n_{s-s'}^{(1)} \rangle_{n',j'}^{\vphantom{(1)}}
    \, \phi_{n'}^{j'}(s') \,,
\end{split}
\end{equation}
where the $\phi_{n'}^{j'}(s')$ are the components of the solution of
the master equations at time $s'$, the $\Phi(n,s,j|n',s',j')$ are the
conditional probabilities as in Eq.~\eqref{pcond1} with $p=2$, and
$\langle n^{(1)}_{s-s'} \rangle_{n',j'}^{\vphantom{(1)}} = \sum_{n \geq 0}
n \, \Phi(n,s,j|n',s',j')$ is the mean mRNA number at time $s$ starting out
with $n'$ mRNA molecules and in promoter state $j'$ at time $s'$. \linebreak
Now the latter is obtained directly by adapting Eq.~\eqref{ncopy} to this
shifted initial time and these initial conditions, resulting in
\begin{equation} \label{nss}
 \langle n^{(1)}_{s-s'} \rangle_{n',j'}^{\vphantom{(1)}}
 = \bar{N} + (n' - \bar{N}) \, \ee^{-(s-s')}
   + \frac{\Delta N}{1-\epsilon} (\delta_{j',1} - p_1) \,
     (\ee^{-\epsilon(s-s')} - \ee^{-(s-s')}) \,,
\end{equation}
where $\delta$ is the Kronecker symbol ($\delta_{j',j}$=1 when $j'=j$
and $\delta_{j',j}=0$ when $j' \neq j$).
From Eqs~\eqref{nsns} and~\eqref{nss}, it follows that, for $s>s'$,
\begin{equation}\label{nsnsp}
\begin{split}
 &\langle n_s n_{s'} \rangle
  - \langle n^{(1)}_s \rangle \langle n^{(1)}_{s'} \rangle \\
 &= V(n_{s'}) \, \ee^{-(s-s')} + \frac{\Delta N}{1-\epsilon} \,
    \bigl( \langle n^{{(1)}}_{s'} \rangle_1^{} - \pi_1(s')
    \langle n^{{(1)}}_{s'} \rangle \bigr) \,
    (\ee^{-\epsilon(s-s')}- \ee^{-(s-s')}) \,.
\end{split}
\end{equation}
From Eqs~\eqref{solsysocpr} \eqref{ncopy} and~\eqref{n1mrna}, it follows
that the quantity $\langle n^{(1)}_s \rangle_1^{\vphantom{1}} - \pi_1(s)
\langle n^{(1)}_s \rangle$ has the structure:
\begin{equation}\label{n1s}
 \langle n^{(1)}_s \rangle_1^{} - \pi_1(s) \langle n^{(1)}_s \rangle
 = A_2 + B_2 \ee^{-\epsilon s} + C_2 \ee^{-(1+\epsilon)s}
   + D_2 \ee^{-2\epsilon s},
\end{equation}
with coefficients:
\begin{equation}
 \begin{array}{c}
 {\displaystyle A_2 = \frac{\Delta N \, p_1 (1-p_1)}{1+\epsilon} \,,} \\[3ex]
 {\displaystyle B_2 = \Delta N \, (1-2p_1) (\pi_1(0)-p_1) \,,} \\[3ex]
 {\displaystyle C_2 = \frac{\epsilon \, \Delta N \,
                            [2\pi_1(0) + \epsilon (1-2p_1) - 1] \,
                            (\pi_1(0)-p_1)}{(1+\epsilon)(1-\epsilon)} \,,}
 \\[3ex]
 {\displaystyle D_2 = - \, \frac{\Delta N \, (\pi_1(0)-p_1)^2}{1-\epsilon} \,.}
 \end{array}
\end{equation}
Using  \eqref{nsnsp},\eqref{vns},\eqref{n1s}, we find, for $s>s'$
\begin{equation}\label{fsum}
 \langle n_s n_{s'} \rangle
 - \langle n^{(1)}_s \rangle \langle n^{(1)}_{s'} \rangle
 = \sum_i K_i \, \ee^{c_i s + d_i s'} \,,
\end{equation}
and similarly, for $s'>s$,
\begin{equation}\label{fsump}
 \langle n_s n_{s'} \rangle
 - \langle n^{(1)}_s \rangle \langle n^{(1)}_{s'} \rangle
 = \sum_i K_i \, \ee^{c_i s' + d_i s} \,,
\end{equation}
with coefficients $K_i$, $c_i$, $d_i$ given in Table~\ref{table1}.

\begin{table}[!h]
\caption{\label{table1} Coefficients in Eqs~\eqref{fsum} and~\eqref{fsump}.}
\begin{center}
\begin{tabular}{|c|r|r|r|}
\hline
 $i$  & $c_i$ & $d_i$ & $K_i$ \\ \hline
 $1$  & $-1$ & $1$ & $A_1 - A_2 \Delta N/(1-\epsilon)$ \\
 $2$  & $-\epsilon$ & $\epsilon$ & $A_2 \Delta N/(1-\epsilon)$ \\
 $3$  & $-1$ & $0$ & $B_1$ \\
 $4$  & $-1$ & $-1$ & $C_1$ \\
 $5$  & $-1$ & $1-\epsilon$ & $D_1 - B_2 \Delta N/(1-\epsilon)$ \\
 $6$  & $-1$ & $-\epsilon$ & $E_1 - C_2 \Delta N/(1-\epsilon)$ \\
 $7$  & $-\epsilon$ & $0$ & $B_2 \Delta N/(1-\epsilon)$ \\
 $8$  & $-\epsilon$ & $-1$ & $ C_2 \Delta N/(1-\epsilon)$ \\
 $9$  & $-\epsilon$ & $-\epsilon$ & $ D_2 \Delta N/(1-\epsilon)$ \\
 $10$ & $-1$ & $1-2\epsilon$ & $F_1 - D_2 \Delta N/(1-\epsilon)$  \\ \hline
\end{tabular}
\end{center}
\end{table}

\noindent
Putting everything together, we are now in a position to evaluate
the integral $I_\tau$ in Eq.~\eqref{mvar}: it has the form
\begin{equation}
 \begin{split}
 I_\tau = \;
 &\sum_i K_i \int_0^\tau \left( \int_{s'}^\tau \ee^{c_i s' + d_i s} \,
  \ee^{\alpha(s+s')} ds \right) ds' \\
 &+ \sum_i K_i \int_0^\tau \left( \int_s^\tau \ee^{c_i s + d_i s'} \,
    \ee^{\alpha(s+s')} ds' \right) ds \,,
 \end{split}
\end{equation}
so evaluating these integrals we get
\begin{equation}
 I_\tau
 = \sum_i \left[ \frac{2 K_i \, \ee^{(2\alpha+c_i+d_i)\tau}}%
                      {(\alpha+d_i)(2\alpha+c_i+d_i)} -
                 \frac{2 K_i \, \ee^{(\alpha+c_i)\tau}}%
                      {(\alpha+c_i)(\alpha+d_i)} +
                 \frac{2 K_{i}}{(\alpha+c_i)(2\alpha+c_i+d_i)} \right].
\end{equation}
This gives us our final result for the protein number density variance:
\begin{equation}
 \begin{split}
 V(m_\tau) = \;
 & V(m_0) \ee^{-2\alpha\tau}
   + \sum_i \frac{2\beta^2 K_i \,
     \ee^{(c_i+d_i)\tau}}{(\alpha+d_i)(2\alpha+c_i+d_i)} \\
 & - \sum_i \frac{2\beta^2 K_i \,
     \ee^{(c_i-\alpha)\tau}}{(\alpha+c_i)(\alpha+d_i)}
   + \sum_i \frac{2\beta^2 K_i \,
     \ee^{-2\alpha\tau}}{(\alpha+c_i)(2\alpha+c_i+d_i)} \,,
\end{split}
\end{equation}
with asymptotic value
\begin{equation} \label{vss}
 \begin{split}
 \lim_{\tau \to \infty} V(m_\tau) \;
 &= \, \frac{\beta^2}{\alpha} \sum_{c_i+d_i=0} \frac{K_i}{\alpha+d_i} \\
 &= \, \frac{\beta^2}{\alpha}
       \left[\frac{A_1}{\alpha+1} + \frac{A_2 \, \Delta N}{1-\epsilon}
             \left(\frac{1}{\alpha+\epsilon} - \frac{1}{\alpha+1} \right) \right] \\
 & =\, \frac{\beta^2}{\alpha(\alpha+1)}
       \left[ \bar N + (\Delta N)^2 \frac{(\alpha+\epsilon +1) \, p_1 (1-p_1)}%
        {(\alpha+\epsilon)(\epsilon+1)} \right].
\end{split}
\end{equation}
The expression~\eqref{vss} can be compared to the steady state protein number
variance obtained from the completely discrete protein expression model in~%
\cite{Innocentini2007}. 
In that model protein number is treated as a discrete variable, whereas in the
present model it is a continuous variable (density). Consequently, we expect to
lose the contribution to the total variance that stems from discreteness of the
protein degradation process. And indeed, our expression~\eqref{vss} lacks
the term $\langle m_\infty \rangle$, as compared to the steady state variance
computed in~\cite{Innocentini2007}.
This term corresponds to the poissonian component added to the protein
variations by the stochastic protein degradation and is negligible with respect
to the total variance when $\alpha \ll 1$, i.e., when the lifetime of the protein
is much larger than the lifetime  of the mRNA.

\section{Results}
\indent

Following the approach discussed above we have calculated the time
dependent probability distributions for mRNA molecules and protein density. 
More precisely, the dynamics of the probability distribution for the
mRNA population is obtained by applying Eq.~\eqref{dgen} to the
exact solution of the master equations~\eqref{sol}, written in terms
of the original variables $t$ and $z$. From~that, we can compute, at
each instant of time $\tau$, the push-forward measure under the
mapping given by Eq.~\eqref{eq:RDESOLAPPROX}, as defined by
Eq.~\eqref{probproduct}.

\begin{figure}[!ht]
\centerline{\includegraphics[width=4.8in,angle=0]{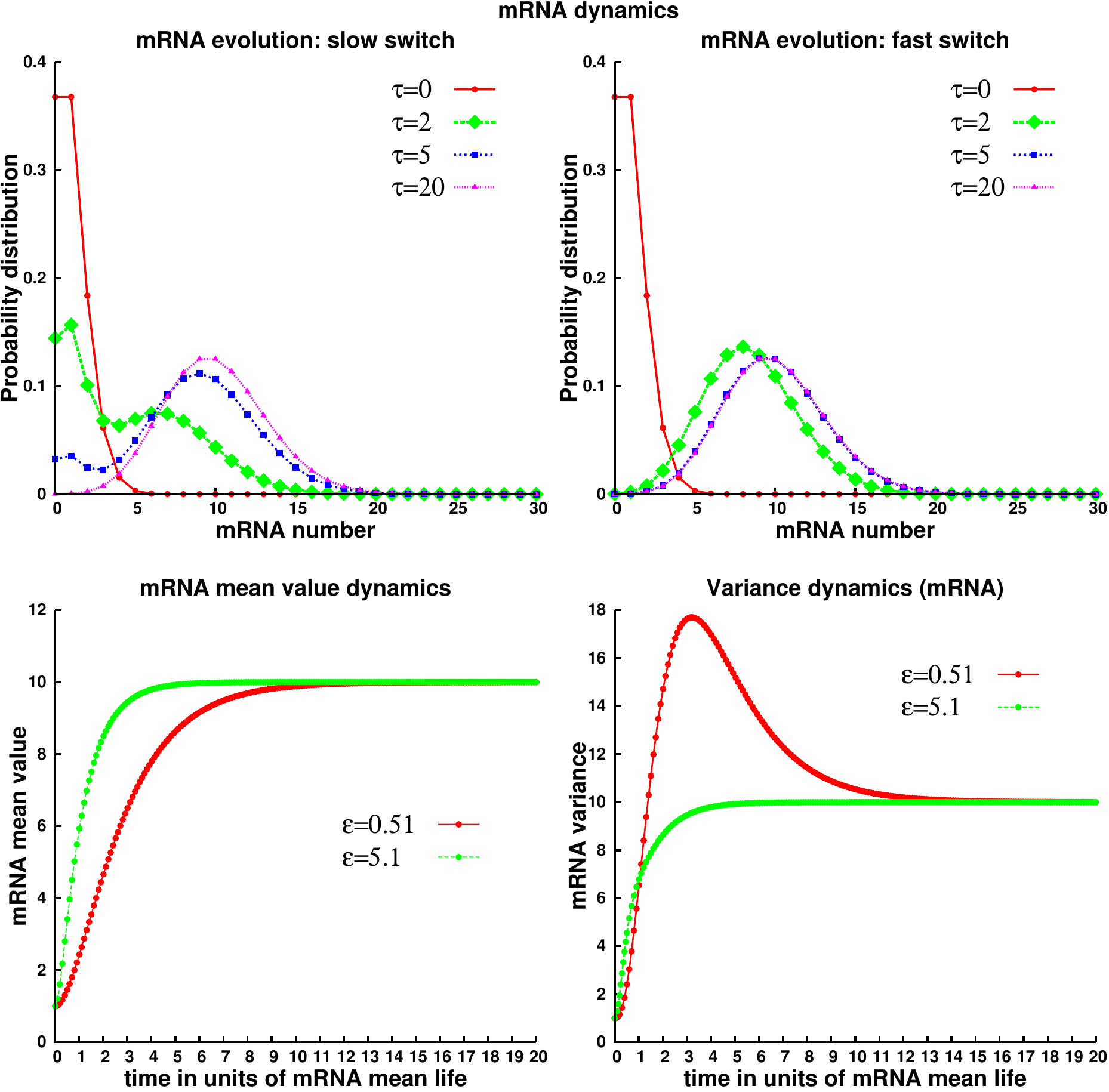}}
\caption{\label{fig1} mRNA dynamics in slow ($\epsilon=0.51$)
and fast ($\epsilon=5.1$) switch regimes. Remaining parameters:
 $N_1=10$, $N_2=1$, $p_1=1$, $p_2=0$.}
\end{figure}

\begin{figure}[!ht]
\subfigure[]{ 
\includegraphics[width=2.5in,height=1.5in,angle=0]{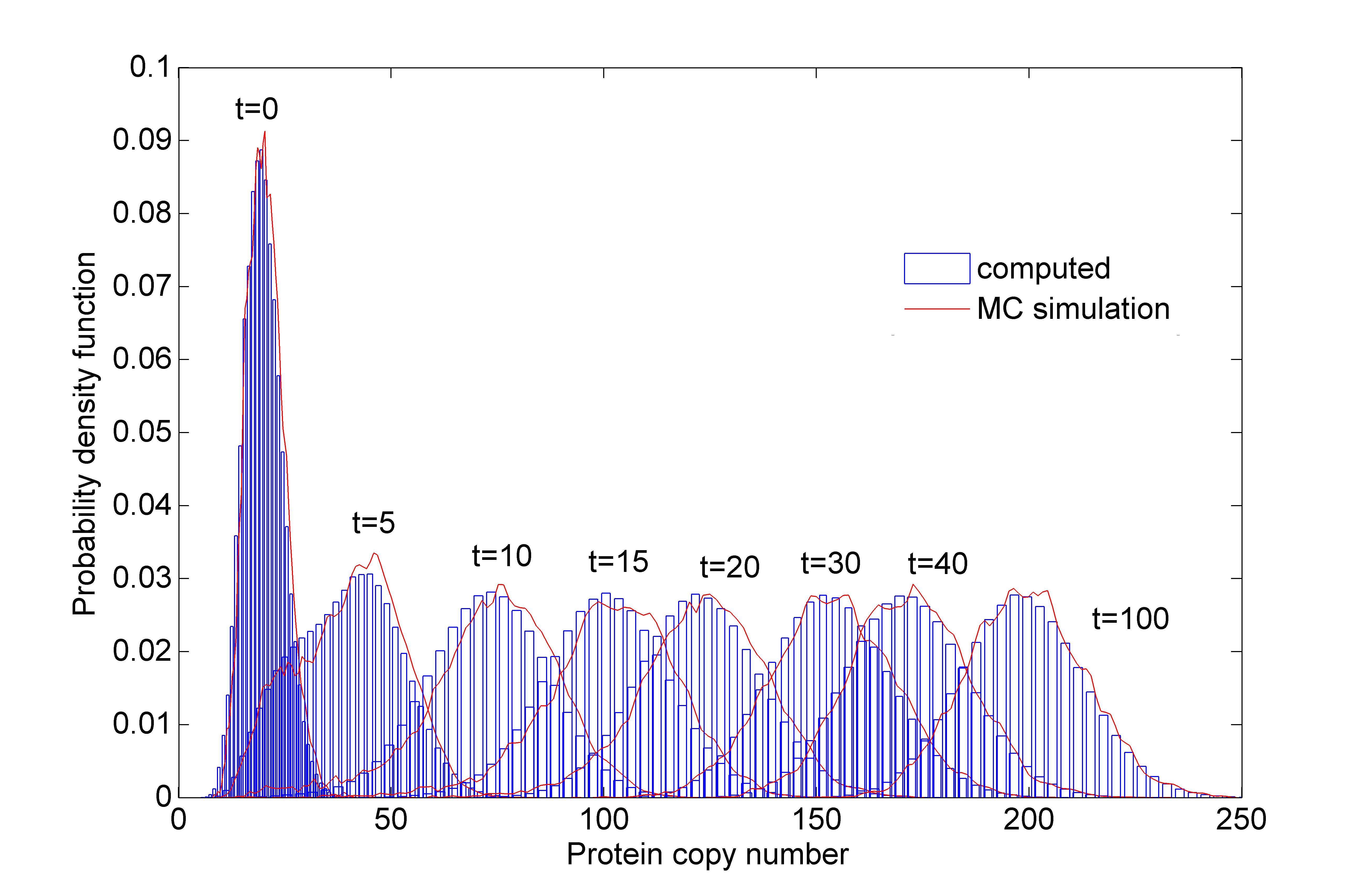}\label{fig2a}}\hspace{-5mm}
\subfigure[]{ 
\includegraphics[width=2.5in,height=1.5in,angle=0]{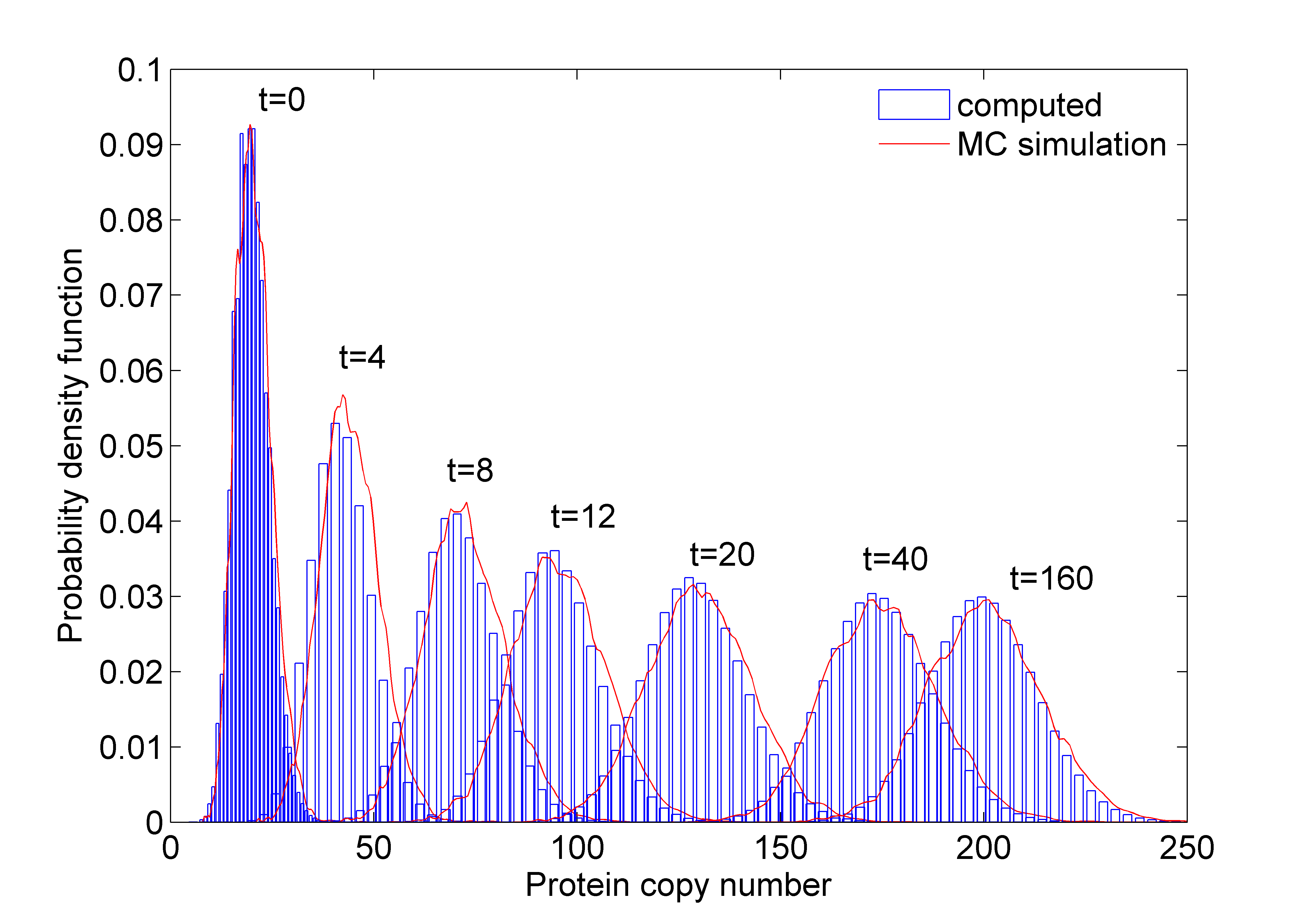}\label{fig2b}}\hfill
\caption{\label{fig2} Comparison with Monte Carlo (MC) simulation.
Slow switch ($\epsilon=0.51$) in Fig.~\ref{fig2a} and fast switch
($\epsilon=5.1$) in Fig.~\ref{fig2b}. Remaining parameters:
$N_1=10$, $N_2=1$, $p_1=1$, $p_2=0$, $\alpha=1/20$, $\beta=1$.}
\end{figure} 

The result of this calculation is an ensemble of protein density values
with their corresponding probabilities, $\{(m_{\tau}^k,\Prob(m_\tau^k)):
k=1,\ldots,(L+1)^p\}$.
Graphically, such ensembles will be represented by histograms where
the probabilities are summed up within each bin. More precisely, if we fix a
bin size and group together all $m_\tau^k$ belonging to the same bin, the
probability assigned to that bin is simply the sum of all the probabilities
$\Prob(m_\tau^k)$ corresponding to the $m_{\tau}^k$ in that bin.

In order to  estimate the accuracy of our method, we compare the
distributions  obtained by our formalism with those from a Monte Carlo
(MC) simulation of the model. We have used the MC simulation to
generate trajectories of the mRNA process $n_\tau$. Namely, let the
$\tau_q$ be the random times when the birth and death process for
mRNA molecules produces a change from $n_\tau$ to $n_\tau \pm 1$.
Then Eq.~\eqref{eq:RDESOLAPPROX} can be used to directly compute
samples of the protein process $m_\tau$. Note that this makes our hybrid
model much easier to simulate than the full discrete mRNA/protein model,
since we avoid the separate simulation of the protein process, which is
computationally costly.

\begin{figure}[!ht]
\subfigure[]{ 
\includegraphics[width=2.5in,angle=0]{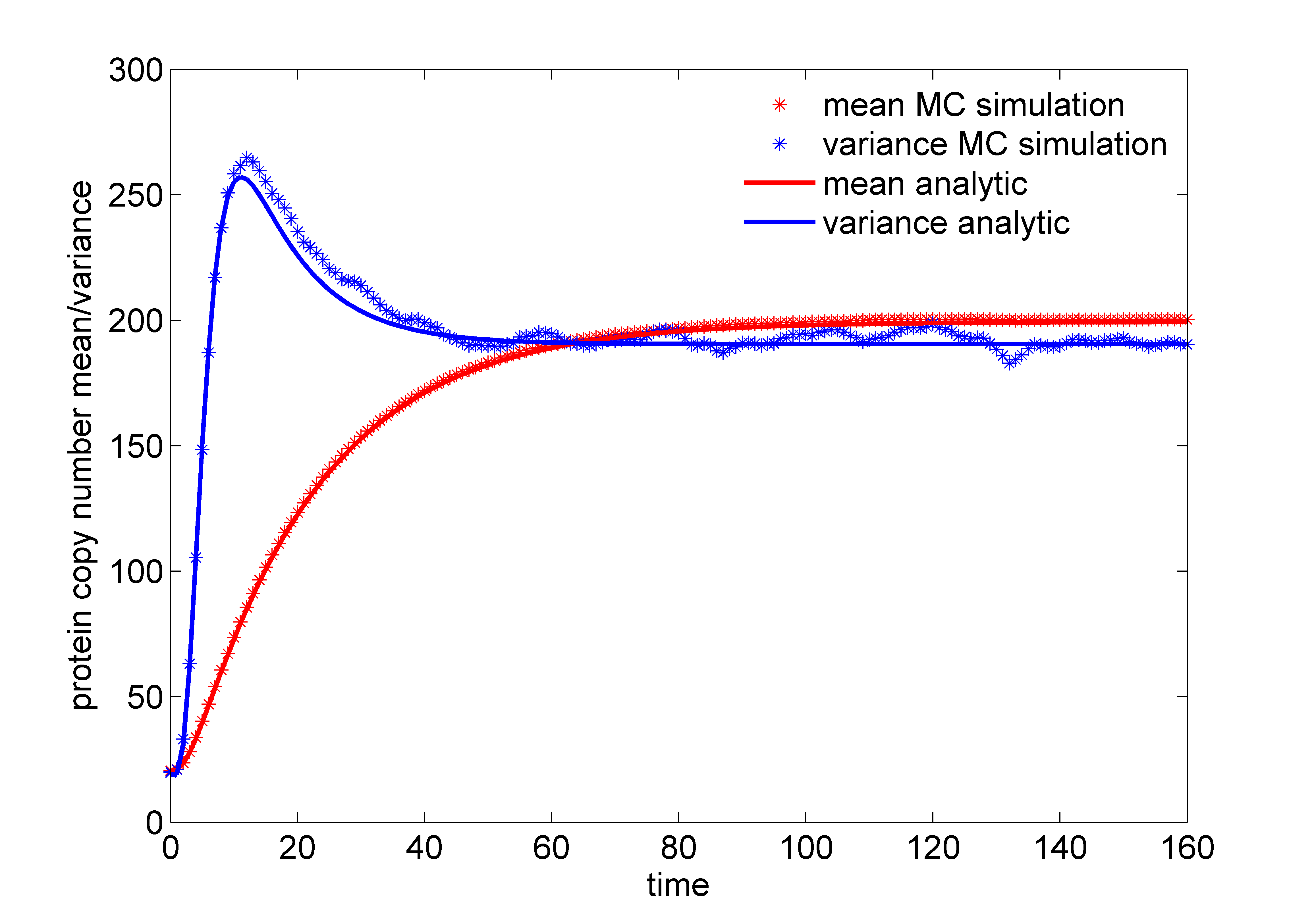}\label{fig3a}}\hspace{-5mm}
\subfigure[]{ 
\includegraphics[width=2.5in,angle=0]{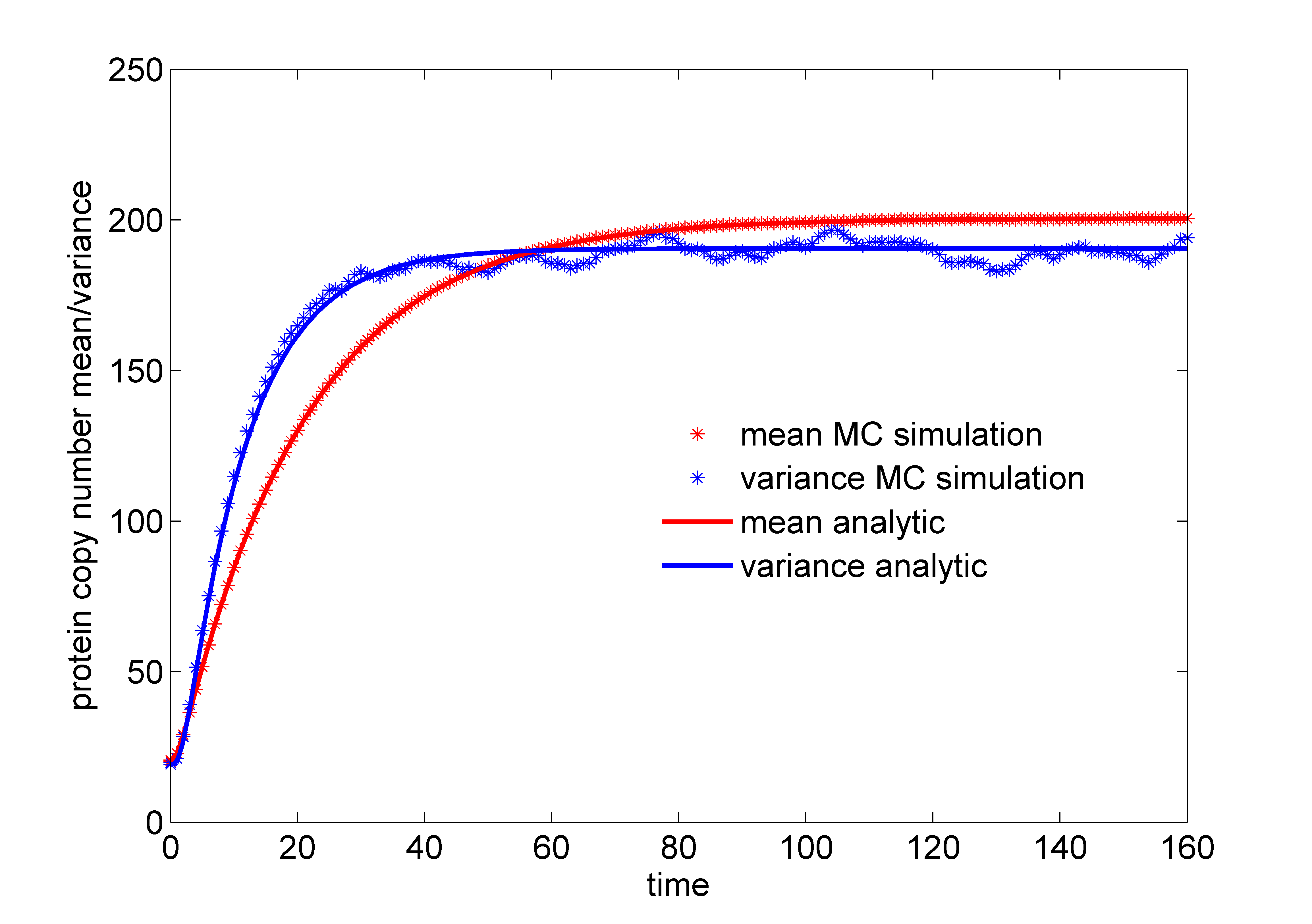}\label{fig3b}}\hfill
\caption{\label{fig3} Dynamical evolution of protein mean value and variance.
Slow switch ($\epsilon=0.51$) in Fig.~\ref{fig3a} and fast switch ($\epsilon=5.1$) in Fig.~\ref{fig3b}. Remaining parameters: $N_{1}=10$, $N_{2}=1$, $p_{1}=1$, $p_{2}=0$, $\alpha=1/20$, $\beta=1$.}
\end{figure}

As an example of our results we exhibit in Fig.~\ref{fig1} 
the time evolution for the probability distribution of the mRNA
population, its mean value and variance.
In all cases, we have used as initial mRNA configuration the generating
function $\, \phi(\mu) = \exp(-N_{2}\mu)$, representing the gene with
probability one in the off state, that is, the initial mRNA number
follows a Poisson distribution with mean equal to $N_2$. 
On the other hand, the occupancy probabilities have been chosen
as $p_1=1$, $p_2=0$, so as to produce a final equilibrium state
which represents the gene in full activity and with mRNA number
following a Poisson distribution with mean equal to~$N_1$. 
Concerning the switching parameter~$\epsilon$, we have selected
two values: $\epsilon=5.1$ representing the ``fast switch regime''
and $\epsilon=0.51$ representing the ``slow switch regime''.
Specifically, in Fig.~\ref{fig2} we exhibit, for the two switch
regimes, a direct comparison between the distributions obtained
by our method (blue histograms) and those from MC simulation
(red curves), and finally, in Fig.~\ref{fig3} we show the mean
value and variance of the protein distribution, comparing the
analytical formulas presented in Section 4 with the results of
a direct simulation of the model.

The transient behavior of mRNA in the slow switch regime has a two peak
distribution, indicating a more noisy configuration as compared to the
fast switch regime, where the distribution is unimodal. 
The multi-modality in the slow switch regime is reflected in the 
protein probability density, where for time $\tau=5$, in Fig.~\ref{fig2a},
one can see the existence of a strong asymmetry. Also, it accounts for
an increase in the noise of protein synthesis in the transient time,
captured in the overshoot in Fig.\ref{fig3a}.
Increasing the gene switch parameter decreases the standard deviation
in mRNA production, which is a well known effect~\cite{Innocentini2007,%
Innocentini2013}.

\section{Discussion and Conclusion}
\indent

The hybrid model presented here shows how to couple transcription and
translation providing a complete picture of the entire dynamical process,
without any restrictions on the parameter space.
The randomness of protein synthesis due to the stochastic nature of
transcription is exhibited in the dynamical behavior of the protein
probability density. 
The main result is a full time-dependent solution for the probability
distribution of mRNA as well as for the density probability for protein
numbers~-- something that, to the best of our knowledge, has never
been achieved before. Moroever, the distributions for protein
number obtained by our method are in excellent agreement with those
derived from MC simulations, at highly reduced computational cost.
But there is a technical issue that must still be overcome.
Namely, in order to improve the precision of our method, we must
use joint probabilities with many events overlapping in a specific
time interval (bigger values of $p$). The optimization in the
implementation of our method necessary to deal with this
issue will be left to future work.   

It is worth mentioning that pure random differential equations (RDE)
models~-- where the processes of mRNA production and of protein 
production are treated on equal footing, using random differential
equations for both~-- have been introduced in~\cite{LPCBK2006} for
the continuous time case and in~\cite{FBB2009,FBA2013} for the
discrete time case.
Similarly, pure master equations (ME) models~-- where the processes
of mRNA production and of protein production are also treated on equal
footing, but using master euations for both~-- have been discussed in
the literature before; see, for instance~\cite{Innocentini2007,Swain2008}.
Both of these approaches are highly interesting and logically perfectly
consistent, but a closer look reveals some drawbacks.
On the one hand, using pure RDE models means that mRNA is represented
by a continuous random variable, which is problematic since the number
of mRNA molecules is small, of the order of a few dozen per gene.
On the other hand, pure ME models are hard to solve explicitly and
one has to resort to simulations or appeal to some approximation scheme 
in order to simplify the equations and then find expressions for the
protein distribution (a discrete probability distribution) that solve
these simplified equations, rather than the original ones.

Recent experiments allowing real-time observation of the expression of 
stochastic protein synthesis in living \emph{Escherichia coli} 
or \emph{Bacillus subtilis} cells, with single molecule
sensitivity~\cite{CFX2006,ovidiu1}, have shown that information about
key parameters of protein expression can be extracted from the steady
state distribution.
Furthermore, measurements of protein concentration can be integrated
with mRNA tagging techniques,  such as MS2, that monitor mRNA production.
The model discussed here can be used to extract quantitative information on
transcription and translation processes from measured mRNA and protein
distributions. In addition, the ability to compute the shape of the protein
distribution may be used to improve the understanding of stochasticity in
biological decision making processes.

Future research will also be dedicated to developing the model
to include other phenomenological aspects of gene expression.
One modification consists in allowing the protein synthesis/degradation
rates to be random variables, thus taking into account the inherent
noise due to the translational process.
The model can also be extended to study eukaryotes, which requires
introducing a time-delay accounting for the transport of mRNA from
the nucleus to the cytoplasm.
Another modification amounts to adding a non-linear term to the RDE,
reflecting a decrease in protein number due to other effects than just
degradation, such as complex formation by dimerization: this will
introduce a bifurcation parameter and ultimately implement the
observed multi-stability in the steady state of protein population
(the bifurcation theory for RDEs can be found in~\cite{Arnold1998}).
In contrast to multi-stability, the multi-modality originating in 
the controlling mechanism of protein synthesis, at the translational
level, can be introduced by allowing the parameter $B$ (or $\beta$)
in Eq.~\eqref{eq:RDE} to be a matrix, turning the RDE for protein
density into a vector equation. The entries of this matrix will 
encode the different levels of translational efficiency. 

Finally, the model can be used as a building block for constructing
mathematical models of gene regulatory networks. More concretely,
the idea is to take several copies of our model and couple them by
allowing the binding/unbinding rates controlling the on/off switch
of any gene to become functions of the mean values of the proteins
expressed by the other genes. Traditionally, this coupling is
performed through Hill type functions which convert protein
densities into binding/unbinding rates. 
This strategy is in accordance with the ubiquitous idea in physics
that simple models serve as building blocks for more complicated ones.

\paragraph{Acknowledgments.}
We would like to thank the referees for their insights.
Work supported by FAPESP, SP, Brazil (G.I., contract 2012/04723-4)
and CNPq, Brazil (G.I., contract 202238/2014-8; M.F., contract
307238/2011-3; F.A., contract 306362/2012-0). O.R. thanks CNRS
and LABEX Epigenmed for support. 



\begin{thebibliography}{10}
\providecommand{\url}[1]{{#1}}
\providecommand{\urlprefix}{URL }
\expandafter\ifx\csname urlstyle\endcsname\relax
  \providecommand{\doi}[1]{DOI~\discretionary{}{}{}#1}\else
  \providecommand{\doi}{DOI~\discretionary{}{}{}\begingroup
  \urlstyle{rm}\Url}\fi

\bibitem{AandS}
Abramowitz, M., Stegun, I.A.: Handbook of mathematical functions with formulas,
  graphs and mathematical tables.
\newblock U.S. Government Printing Office (1964)

\bibitem{Arnold1998}
Arnold, L.: Random dynamical systems.
\newblock Springer-Verlag, Berlin (1998)

\bibitem{BKCC2003}
Blake, W.J., Kaern, M., Cantor, C.R., Collins, J.J.: Noise in eukaryotic gene
  expression.
\newblock Nature \textbf{422}, 633--637 (2003)

\bibitem{CFX2006}
Cai, L., Friedman, N., Xie, X.: Stochastic protein expression in individual
  cells at the single molecule level.
\newblock Nature \textbf{440}(7082), 358--362 (2006).
\newblock \doi{10.1038/nature04599}

\bibitem{CT1981}
Cogburn, R., Torrez, W.C.: Birth and death processes with random environments
  in continuous time.
\newblock J Appl Probab \textbf{18}(1), 19--30 (1981)

\bibitem{delbruck}
Delbr{\"u}ck, M.: Statistical fluctuations in autocatalytic reactions.
\newblock Journal of Chemical Physics \textbf{8}, 120--124 (1940)

\bibitem{Elowitz2002}
Elowitz, M.B., Levine, A.J., Siggia, E.D., Swain, P.S.: Stochastic gene
  expression in a single cell.
\newblock Science \textbf{297}(5584), 1183--1186 (2002).
\newblock \doi{10.1126/science.1070919}

\bibitem{ovidiu1}
Ferguson, M., Le~Coq, D., Jules, M., Aymerich, S., Radulescu, O., Declerck, N.,
  Royer, C.: Reconciling molecular regulatory mechanisms with noise patterns of
  bacterial metabolic promoters in induced and repressed states.
\newblock P Natl Acad Sci USA \textbf{109}(1), 155--160 (2012)

\bibitem{FBB2009}
Ferreira, R.C., Bosco, F.A.R., Briones, M.R.S.: Scaling properties of
  transcription profiles in gene networks.
\newblock Int J Bioinform Res Appl \textbf{5}(2), 178--186 (2009)

\bibitem{FBA2013}
Ferreira, R.C., Briones, M.R.S., Antoneli, F.: A model of gene expression based
  on random dynamical systems reveals modularity properties of gene regulatory
  networks.
\newblock Preprint, arXiv:1309.0765

\bibitem{Friedman2006}
Friedman, N., Cai, L., Xie, X.S.: Linking stochastic dynamics to population
  distribution: an analytical framework of gene expression.
\newblock Phys Rev Lett \textbf{97}(16), 168302 (2006).
\newblock \doi{10.1103/PhysRevLett.97.168302}

\bibitem{GPZC2005}
Golding, I., Paulsson, J., Zawilski, S., Cox, E.: Real-time kinetics of gene
  activity in individual bacteria.
\newblock Cell \textbf{123}(6), 1025--1036 (2005).
\newblock \doi{10.1016/j.cell.2005.09.031}

\bibitem{Hornos2005}
Hornos, J.E.M., Schultz, D., Innocentini, G.C.P., Wang, J., Walczak, A.M.,
  Onuchic, J.N., Wolynes, P.G.: Self-regulating gene: An exact solution.
\newblock Phys Rev E \textbf{72}(5), 051907 (2005).
\newblock \doi{10.1103/PhysRevE.72.051907}

\bibitem{Innocentini2013}
Innocentini, G.C.P., Forger, M., Ramos, A., Radulescu, O., Hornos, J.E.M.:
  Multimodality and flexibility of stochasctic gene expression.
\newblock Bull Math Biol \textbf{75}, 2600--2630 (2013)
\newblock \doi{10.1007/s11538-013-9909-3}

\bibitem{Innocentini2007}
Innocentini, G.C.P., Hornos, J.E.M.: Modeling stochastic gene expression under
  repression.
\newblock Journal of Mathematical Biology \textbf{55}(3), 413--431 (2007).
\newblock \doi{10.1007/s00285-007-0090-x}

\bibitem{Jayaprakash2009}
Iyer-Biswas, S., Hayot, F., Jayaprakash, C.: Stochasticity of gene products
  from transcriptional pulsing.
\newblock Phys Rev E \textbf{79}(3), 031911 (2009).
\newblock \doi{10.1103/PhysRevE.79.031911}

\bibitem{vKampen2007}
van Kampen, N.G.: Stochastic Processes in Physics and Chemistry, 3rd edn.
\newblock Elsevier, Amsterdam (2007)

\bibitem{KE2001}
Kepler, T.B., Elston, T.C.: Stochasticity in transcriptional regulation:
  origins, consequences, and mathematical representations.
\newblock Biophys J \textbf{81}(6), 3116--3136 (2001).
\newblock \doi{10.1016/S0006-3495(01)75949-8}

\bibitem{LPCBK2006}
Lipniacki, T., Paszek, P., {Marciniak-Czochra}, A., Brasier, A.R., Kimmel, M.:
  Transcriptional stochasticity in gene expression.
\newblock J Theor Biol \textbf{238}, 348--367 (2006).
\newblock \doi{10.1016/j.jtbi.2005.05.032}

\bibitem{OTKGO2002}
Ozbudak, E.M., Thattai, M., Kurtser, I., Grossman, A.D., {van Oudenaarden}, A.:
  Regulation of noise in the expression of a single gene.
\newblock Nature Genetics \textbf{31}, 69--73 (2002)

\bibitem{Paulsson2005}
Paulsson, J.: Models of stochastic gene expression.
\newblock Phys Life Rev \textbf{2}, 157--175 (2005)

\bibitem{PY1995}
Peccoud, J., Ycart, B.: Markovian modeling of gene product synthesis.
\newblock Theor Popul Biol \textbf{48}(2), 222--234 (1995).
\newblock \doi{10.1006/tpbi.1995.1027}

\bibitem{PE2004}
Pirone, J., Elston, T.: Fluctuations in transcription factor binding can
  explain the graded and binary responses observed in inducible gene
  expression.
\newblock J Theor Biol \textbf{226}, 111--121 (2004).
\newblock \doi{10.1016/j.jtbi.2003.08.008}

\bibitem{Ramos2011}
Ramos, A.F., Innocentini, G.C.P., Hornos, J.E.M.: Exact time-dependent
  solutions for a self-regulating gene.
\newblock Phys Rev E \textbf{83}(6), 062902 (2011).
\newblock \doi{10.1103/PhysRevE.83.062902}

\bibitem{ROS2004}
Raser, J.M., O'Shea, E.K.: Control of stochasticity in eukaryotic gene
  expression.
\newblock Science \textbf{304}(5678), 1811--1814 (2004).
\newblock \doi{10.1126/science.1098641}

\bibitem{Swain2008}
Shahrezaei, V., Swain, P.S.: Analytical distributions for stochastic gene expression.
\newblock P Natl Acad Sci USA \textbf{105}(45), 17256--17261 (2008).
\newblock \doi{10.1073/pnas.0803850105}.

\bibitem{YXRLX2006}
Yu, J., Xiao, J., Ren, X., Lao, K., Xie, X.S.: Probing gene expression in live
  cells, one protein molecule at a time.
\newblock Science \textbf{311}(5767), 1600--1603 (2006).
\newblock \doi{10.1126/science.1119623}

\end{thebibliography}

\end{document}